# ALGORITHM FOR SPATIAL CLUSTERING WITH OBSTACLES

Mohamed E. El-Sharkawi
Dept. of Information Systems
Faculty of Computers & Information
Cairo University
Cairo, Egypt
mel_sharkawi@hotmail.com

Mohamed A. El-Zawawy
Dept. of Mathematics
Faculty of Science
Cairo University
Cairo, Egypt
mzawawy@operamail.com

## Abstract

*In this paper, we propose an efficient clustering technique to solve the problem of clustering in the presence of obstacles. The proposed algorithm divides the spatial area into rectangular cells. Each cell is associated with statistical information that enables us to label the cell as dense or non-dense. We also label each cell as obstructed (i.e. intersects any obstacle) or non-obstructed. Then the algorithm finds the regions (clusters) of connected, dense, non-obstructed cells. Finally, the algorithm finds a center for each such region and returns those centers as centers of the relatively dense regions (clusters) in the spatial area.*

1. **Introduction**

Spatial databases contain spatial-related information such databases include geographic (map) databases, VLSI chip design databases, and medical and satellite image databases. Spatial data mining is the discovery of interesting characteristics and patterns that may exist in large spatial databases. It can be used in many applications such as seismology (grouping earthquakes clustered along seismic faults), minefield detection (grouping mines in a minefield), and astronomy (grouping stars in galaxies). Clustering, in spatial data mining, is a useful technique for grouping a set of objects into classes or clusters such that objects within a cluster have high similarity among each other, but are dissimilar to objects in other clusters.

In the last few years, many effective and scalable clustering methods have been developed. These methods can be categorized into partitioning methods [KR90, NH94, BFR98], hierarchal methods [KR90, ZRL96, GRS98, KHK98], density-based methods [EKSX96, ABKS99, HK98], grid-based methods [WYM97, SCZ98, AGG98], model-based methods [SD90, Koh82], and constrained-based methods [THH01]. Most of these algorithms, however, provide very few avenues for users to specify real life constraints such as physical obstacles.

In many applications, physical obstacles like mountains and rivers could substantially affect the result of a clustering algorithm. For example, consider a telephone-company planer who wishes to locate a suitable number of telephone cabinets in the area shown in Figure 1(a) to serve the customers who are represented by points in the figure. In such a situation, however, natural obstacles exist in the area and they should not be ignored. Ignoring these obstacles will result in clusters like those in Figure 1(b), which are obviously inappropriate. For example, a river splits cluster $cl_1$, and some customers on one side of the river will have to travel a long way to reach the telephone cabinet at the other side. Thus the ability to handle such real life constraints in a clustering algorithm is very important.

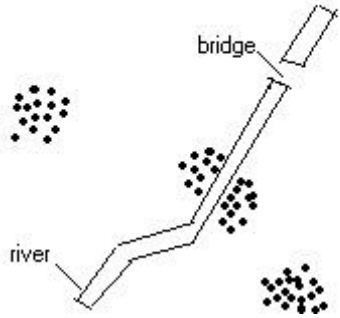 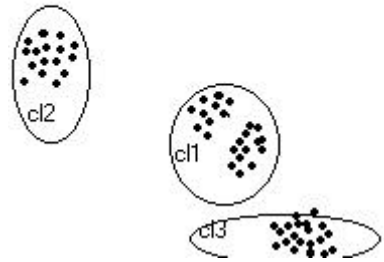

(a) Customers' locations and obstacles            (b) Clusters formed when ignoring obstacles

Figure 1. Planning the locations of ATMs

In this paper, we propose an efficient spatial clustering technique, *SCPO*, which considers the presence of obstacles. The algorithm divides the spatial area into rectangular cells and labels each cell as either *dense* or *non-dense* (according to the number of points in this cell) and also as either *obstructed* (intersects any obstacle) or *non-obstructed.* The algorithm finds all the *dense, non-obstructed* regions that form the clusters by a breadth-first search and determines a center for each region(cluster).

The proposed algorithm has the following advantages over the work that has done to solve the problem of clustering in the presence of obstacles [THH01].
1. It handles outliers or noise. Outliers refer to spatial objects, which are not contained in any cluster and should be discarded during the mining process.
2. It dose not use any randomized search.
3. Instead of specifying the number of the desired clusters beforehand, it finds the natural number of clusters in the spatial area.
1    4. When the data is updated, we do not need to re-compute all information in the cell grid. Instead, incremental update can be done.

The rest of the paper is organized as follows. We first discuss the related work in Section 2. The general algorithm is given in Section 3. Section 4 illustrates how to mark the obstructed cells and determine whether a point is inside an obstacle. In Section 5, we analyze the complexity of our algorithm. Section 6, concludes the paper.

2. **Related Work**

Many studies have been conducted in cluster analysis. Methods related to our work are discussed briefly in this section and we emphasize what we believe are limitations which are addressed by our approach.

**CLARANS**

CLARANS (Clustering Large Applications based up on RANdomized Search) was introduced in [NG94] as the first clustering technique in spatial data mining. The algorithm takes as an input the number, $k$, of the desired clusters but such a parameter is often hard to determine. So a good clustering algorithm should minimize the input parameters. First, the algorithm randomly selects $k$ points as the centers for the required clusters and then tries to find better solutions. Better solution means a new set of centers that minimize the summation of the distances that each object has to cover to the center of its cluster. The computational complexity of CLARANS is $O(n^2)$, where $n$ is the number of the objects.

Since the randomized search is used in the algorithm, the quality of the results cannot be guaranteed when N is large. In addition, CLARANS assumes that all objects are stored in main memory. This clearly limits the size of the database to which CLARANS can be applied

**COD-CLARANS**

COD-CLARANS (clustering with *obstructed* distance based on CLARANS) [THH01] was the first clustering algorithm that takes into consideration the presence of obstacle entities. COD-CLARANS contains two phases. The first phase breaks the database into several databases and summarizes them individually by grouping the objects in each sub-database in micro-clusters. A micro-cluster is a group of points, which are so close together that they are likely to belong to the same cluster. The second phase is the clustering stage. The algorithm randomly selects $k$ points as the centers for the required clusters and then tries to find better solutions.

The complexity of the algorithm is $O(m|V|)+O(m^2|V|)+O(N|V|)$. Where $m$ is the number of the micro-clusters to be clustered, $N$ is the number of the data objects in the database $D$, and $V$ is the number of the vertices of the obstacles. Although COD-CLARANS generates good clustering results, there are several major problems with this algorithm.. Since COD-CLARANS is an extension of CLARANS algorithm, it suffers from similar drawbacks as CLARANS. In addition, COD-CLARANS can't handle outliers.

**3. Proposed Algorithm**

In this section, we describe our proposed algorithm, *SCPO* (Spatial Clustering in the Presence of Obstacles), in detail. The algorithm first divides the spatial area into $m$, which is an input parameter, rectangular cells of equal areas. Then, the algorithm labels each cell as *dense* or *non-dense* (according to the number of points in that cell and an input threshold). The algorithm also labels each cell as *obstructed* (i.e. intersects any obstacle) or *non-obstructed*. The algorithm finds maximal connected regions of *dense, non-obstructed* cells. The algorithm marks obstructed cells as follows  Given an obstacle and the minimum, say $e$, of the two dimensions of a cell in the grid structure, the algorithm marks the cells that intersect the boundary of an obstacle as follows. For each boundary edge of the obstacle, the algorithm marks the cells that intersect the vertices of that edge as obstructed. Then the algorithm divides this edge by the mean point of the two vertices that determine it and marks the cell that contains the mean point as obstructed and repeat the same procedure with the two segments that constituted by the mean point until the length of each of them is greater than or equal $e$.

The algorithm finds a center for each obtained region. To find a center for a region the algorithm first determines the arithmetic mean of all the points in the region. If that mean point is not obstructed (lies in an obstructed cell), the algorithm returns it as the center of the region. Otherwise the algorithm calculates a cost for each cell in the region. The cost of a cell is the summation of obstructed distances [THH01] between the mean point of the cell in hand and the mean points of all the other cells in the region with each distance multiplied by the number of points in the far cell. Then the algorithm returns the mean point of the cell with the smallest cost as the center of the region. While we calculate a center for a cluster we need to check whether a point is inside an obstacle. For doing that our algorithm applies the Ray Crossings algorithm [O'R98], which takes $O(t)$ to determine whether a point is inside an obstacle that determined by $t$ vertices.

*Algorithm 3.1 SCPO.*
*Input*:
1. A set of $N$ objects and a set of polygon obstacles $\{O_1,O_2,...,O_t\}$ in a spatial area $S$.
2. A squared number, $m$, which represents the number of cells in the spatial area such that $m<<N$.
3. A percentage, $h$.

*Output*: Clusters  in $S$ along with their centers.
*Method*:
1. $w = \sqrt{m}$;
2. Divide the spatial area into $m$ rectangular cells by dividing each dimension of the spatial area into $w$ equal segments;
3. For($i=0$; $i<m$; $i++$)
   {
   determine for the cell, $c_i$, parameters:
   - $n_{c_i}$ : the number of the objects in that cell.
   - $m_{c_i}$ : the mean object of all objects in that cell.
   }
1. $d = round(\frac{N}{m} * h)$;
2. For($i=0$; $i<m$; $i++$)
   {

if ($n_{c_i} \geq d$) then
                        $c_i$ is labeled as *dense*;
                    else
                        $c_i$ is labeled as *non-dense*;
                }
    3.  For($i=0$; $i<t$; $i++$)
        {
            Label all cells that intersect the boundary of $O_i$ as *obstructed*;
        }
    4.  $j=0$;
    5.  For($i=0$; $i<m$; $i++$)
        {
            if ($c_i$ is *dense* and *non-obstructed*) then,
            {
                if ($c_i$ is not processed yet) then,
                {
                    - Construct a new cluster, $r_j$, and mark the cell $c_i$ as an element of $r_j$;
                    - Put the *dense, non-obstructed* neighboring cells of $c_i$ in a list, $Q$;
                    - While ($Q$ is not empty)
                        {
                            - Take the first element, $c'$, from $Q$, mark $c'$ as an element of the cluster $r_j$ and add to $Q$ the *dense, non-obstructed* neighboring cells of $c'$, that have not been processed yet;
                        }
                    - $j++$;
                }
            }
        }
    6.  For($i=0$; $i<j$; $i++$)
        {
            Find_center($r_j$);
        }
    7.  Output the clusters with their centers;

*Algorithm 3.2 Find_center(r)*
*Input:* A cluster, $r$.
*Output*: a center for the cluster $r$.
*Method:*
    1.  Calculate the mean point, $mp$, of all the points in the region r.
    2.  If ($mp$ is not in any obstacle), then
            return $mp$;
        else
        {
            For($i=0$; $i<m$; $i++$)
                If ($c_i \in r$) then,
                    $$cost(c_i) = \sum_{j \in r} n_j (d'(m_{c_i}, m_j))^2;$$
        }//where $d'(m_c, m_i)$ is the obstructed distance [THH01], between $m_c$ and $m_i$.
    3.  Return the mean point of the cell in $r$ that have the minimum cost.

**4. Complexity Analysis of the *SCPO* Algorithm**

We will analysis the complexity of the *SCPO* in this section. The following notation are defined for this discussion:
    $N$: the number of data objects in the database $D$.
    $m$: the number of the cells in the spatial area, $m<<N$.

G: the visibility graph for the vertices of the obstacles with a set of vertices *V* and a set of edges *E* that connect pair of vertices that are mutually visible.

|*V*|: the number of vertices in *G*.

|*E*|: the number of the edges in *E*.

Steps 1, 2, 4, and 7 take a constant time. In step 3, we need to scan the database only once so step 3 takes $O(N)$ time. In the worst case, steps 5 and 6 each may need to scan all cells. So, the running time for steps 5 and 6 $O(m)+O(m)= O(m)$ time. Step 8 obviously takes $O(m)$ time.

The running time of step 9 is $O(m|V|^2)+O(m^2/|V|)$. This is because in the worst case, which occurs when the mean point of the points in a region is inside an obstacle, we will need to scan all cells in the region and find the cell whose mean point is the nearest to the mean points of all the other cells in the region. For each cell, to determine the summation of the obstructed distances between its mean point and the mean points of the other cells in the region, this takes $O(|V|^2)+O(m/|V|)$ [J99]. Since the calculation of summation is needed for each cell in the region, the total complexity of doing this is $O(m|V|^2)+O(m^2/|V|)$ and the Ray Crossings algorithm[O'R98] takes $O(|V|)$ to determine whether the mean point is inside an obstacle.

The total complexity of *SCPO*-algorithm is the sum of the running time of all the steps:

$$O(SCPO) = O(N) + O(m) + O(m) + O(m|V|^2) + O(m^2/|V|)$$
$$= O(N) + O(m|V|^2) + O(m^2/|V|).$$

## 5. Conclusion

In this paper, we introduced a new approach to spatial clustering in the presence of obstacles. The proposed algorithm, *SCPO*, finds clusters in the spatial area along with their centers taking into consideration existing of natural obstacles, The algorithm has several advantages over existing algorithms. First, it does not use randomized search to determine the center of a cluster, therefore the quality of results is guaranteed. Second the algorithm requires minimum input. Third, outliers are efficiently handled as they are disregarded from the clusters. The computational complexity of the algorithm is comparable to the best known clustering algorithm.

Future work will consider comparing the performance of our algorithm with COD-CLARANS. Extending the algorithm to work for *d* dimensions, *d* >2, is an interesting and challenging task.